Deformation mechanisms in a TiNi shape memory alloy during cyclic loading

Anne-Lise Gloanec [a*], Giovambattista Bilotta [b], Michel Gerland [b]

[a] UME/MS – EnstaParistech – 828 Boulevard des Maréchaux – 91762 Palaiseau Cédex – France
[b] Institut P' – Département Physique et Mécanique des Matériaux – UPR3346 – ENSMA – Téléport 2 – 1 Avenue Clément Ader, BP 40109, 86961 Futuroscope Chasseneuil Cédex – France

[*]*Corresponding Author*: Anne-Lise Gloanec – UME/MS – EnstaParistech – 828 Boulevard des Maréchaux – 91762 Palaiseau Cédex – France; Tel: +33 169319750; Fax: +33 169319906; e-mail: anne-lise.gloanec@ensta-paristech.fr

**Abstract**

Deformation mechanisms governing the cyclic stress-strain behaviour of a TiNi shape memory alloy are investigated in this work. In order to understand the implementation of these mechanisms during cyclic loading, three low cycle fatigue tests were performed and stopped at different stages. The first test was stopped after the first cycle, the second one after 40 cycles corresponding to the beginning of the stabilization of the cyclic strain-stress behaviour and the last one was carried out up to failure (3324 cycles). Submitted to fatigue loading, the response of the TiNi shape memory alloy presents a classical pseudoelastic response. Two deformation mechanisms, identified by TEM observations, are highlighted, the first one by twins and the second by dislocation slip and their interaction with precipitates. These two mechanisms evolve without competition during cyclic loading. In the same time, nanomechanical properties were also examined and an evolution of the microhardness or indentation modulus has been noticed.

**1 – Introduction**

Shape Memory Alloys (SMAs) are widely applied in various industrial fields such as aeronautic, biomedical, automobile, civil engineering or domestic [1, 2]. TiNi alloys are popular SMAs due to their excellent biocompatibility and mechanical properties, including low weight, good corrosion resistance, and good fatigue strength [3, 4]. SMAs are fascinating materials which present properties that usually do not characterize ordinary metals and alloys. Indeed, TiNi shape memory alloys exhibit very specific thermomechanical behaviours, including shape memory effect [3, 5, 6], good superelasticity [7-9] and high damping capacity at room temperature [5, 10, 11]. Many engineering applications have been developed using some of these properties.
Within the frame of design and reliability of systems using SMAs, it is essential to have phenomenological models representing thermomechanical behaviour as closely as possible fatigue. Understanding the physical mechanisms governing the cyclic behaviour and leading to the degradation is a necessary step.

SMAs can undergo limited plastic deformation in a specific temperature range and then quickly revert back to their original shape in their high-temperature state through complete stress relaxation. Such behaviour is called superelasticty, also designated pseudoelasticity. In this paper, we focus on this particular behaviour. The purpose of this study is to establish a relation between cyclic loading and specific properties (transformation temperature, microhardness, indentation modulus), in one hand, and deformation mechanisms, in the other hand.

## 2 – Material and experimental test procedures

### 2.1 – Material

A commercial TiNi shape memory alloy containing 51.3% (at.) Ti was used in this investigation. In order to develop pseudoelastic behaviour, a particular thermomechanical treatment was carried out. The various stages of this thermomechanical treatment were already described elsewhere [12]. The resulting microstructure is then mainly made of fine grains, with an average size of 25 µm.

### 2.2 – Experimental test procedures

The test specimens used were cylindrical with a gauge section of 8 mm in diameter, 20 mm in length and a total length of 120 mm. A mechanical polishing of the gauge length was carried out with silicon carbide paper in order to minimize the effects of the surface irregularities, like work hardening due to machining or oxide layers developed at 850°C, during the first step of the thermo-mechanical treatment. The final surface preparation was then achieved by an electrolytic polishing. Low Cycle Fatigue (LCF) tests were conducted on a servo-hydraulic machine (MTS 810) controlled in force (from 0 to 23 kN or 0 MPa to 485 MPa) at 1Hz. In order to work in the austenitic phase, the tests were performed at 50°C. This temperature was determined by using the results of differential scanning calorimetry. The signal was sinusoidal in shape with a null stress ratio. The strain amplitude was measured by an EPSILON extensometer with a root of 10 mm, placed on the gauge of the test specimen. The objective of this work was to identify the deformation mechanisms which govern the Cyclic-Strain-Stress (CSS) behaviour. So three LCF tests were performed and stopped at different stages. The first one was stopped after the first cycle, the second one after 40 cycles, corresponding to the beginning the CSS behaviour stabilization [12-14]. The last test was performed until failure ($N_f$ = 3324 cycles).

After fatigue tests, samples were cut from the gauge length of the test specimens for Differential Scanning Calorimetry (DSC) measurements, for microhardness tests and for the microstructural characterisation by Transmission Electron Microscopy (TEM).

The first step, in characterizing a SMA material, is to determine the characteristic transformation temperatures. There are several transformation temperatures, including the austenite start temperature ($A_s$), and the austenite finish temperature ($A_f$) during heating and the martensite start temperature ($M_s$) and the martensite finish temperature ($M_f$) during cooling. Additionally, an intermediate phase (R-phase) often appears during cooling, having its own start temperature ($R_s$) and finish temperature ($R_f$), before the transformation proceeds to martensite at lower temperature. All these transformation temperatures can be obtained by DSC measurements [15] with a TA DSC Q20 machine. Initially, the material specimen was cooled from approximately 80°C to - 50°C, and then heated from - 50°C to 80°C. The starting and finishing temperatures of each transition phase ($M_s$, $M_f$, $A_s$, $A_f$, $R_s$ and $R_f$), the peak temperatures ($M_p$, $A_p$ and $R_p$) and the heat flow were determined from the DSC thermogram. Note that several cycles of heating and cooling were carried out to be sure that there was not cycling effect on the transformation temperatures.

The microhardness tests were performed on samples polished to 1 µm with a Fischerscope H100C apparatus equipped with a square base pyramidal indenter. Several squares (200 µm x 200 µm) of 49 indentations were done on each sample with a load of 30 mN and a load application time of 15 s. A creep time of 15 s was also added at the maximum load before unloading. The microhardness, the indentation modulus and the creep behaviour were obtained for each sample.

The cyclic deformation microstructures were examined in a Phillips CM20 transmission electron microscope operated at 200kV. Discs of 3 mm diameter were cut from the gauge length of the bulk specimens near the fracture surface for the test conducted until failure and at the middle of the gauge length for the two others. In each case, discs were cut perpendicularly to the loading axis by means of electric-discharge machining. The thickness of the samples was reduced to about 75 µm by mechanical polishing. Electron transparent foils for TEM were then prepared using the electrolytic double-jet technique with an electrolyte of 95% acetic acid – 5% perchloric acid solution. For each fatigue sample, several thin foils were studied to obtain large observation areas and to have a representative point of view.

## 3 – Results and Discussion

### 3-1 – Cyclic strain-stress behaviour

The cyclic strain-stress behaviour of a TiNi shape memory alloy has already been presented elsewhere [7, 12-14]. Test was performed at 50°C to work in the austenite phase. A classical pseudoelastic response is observed. An example of a characteristic loop is given in Figure 1. The hysteresis loop consists of six distinct segments. From O to A, the response corresponds to the elasticity of austenite. This linear curve is followed, from A to B, by a stress plateau where the transformation austenite to martensite constantly occurs. At point B, the specimen is then completely in martensite phase, and the response, from B to C, corresponds to the elasticity of this last one. Afterwards, the stress is relaxed (point C). From C to D, the behaviour of the martensite phase is still elastic. Reverse transformation, martensite to austenite, starts in point D and continues in the stress plateau, to point E. Finally, the material is completely in austenite phase and the response, from E to O, corresponds to the elasticity of austenite. In our case, at the end of the first cycle, no residual strain was noticed. During cycling, the strain-stress response evolves and reaches a stabilized state. Indeed, the hysteresis loops are modified, changing their form and becoming smaller (Figure 2). Nevertheless, this change tends to stabilize with increasing the number of cycles. This stabilization effect occurs during the first hundred cycles [9]. Here, the stabilization is reached, around 40 cycles.

Another way to present the cyclic behaviour is the dissipated energy versus the number of cycles [7, 12], the dissipated energy being equal to the surface of the hysteresis loop in the strain-stress curve. As reported in Figure 2, the behaviour shows a rapid decrease during the fifty first cycles followed by a slight decrease up to around the 500$^{th}$ cycle and then stabilization until failure.

**3-2 – Differential Scanning Calorimetry**

Transformation temperatures for each cycled sample were reported in the DSC thermogram (heat flow vs. temperature, Figure 3a for tested material and Figure 3b for untested material).

To be sure that there was not cycling effect on the transformation temperatures, for each sample, two cycles of heating and cooling were achieved. The heat flow curves, corresponding to the both cycles on each sample, are quite superimposed with very tiny differences. When it is discernible, the peaks are always slightly higher in the second cycle.

It is a well-known fact that according to temperature TiNi shape memory alloys exhibit two phases: martensite and austenite [16-18]. Martensite transformations are solid-to-solid phase transformations that occur without diffusion or plasticity, potentially making them reversible. They involve changes in crystalline structure that can be induced by changes in either temperature or stress. The high-temperature, stress-free phase is called austenite (A-phase), which has a high-symmetry crystal structure usually based on a cubic lattice (CsCl structure or B2). The low-temperature, stress-free phase is called martensite (M-phase), which has a lower symmetry based on a monoclinic structure (B19'). In some special situations, there is an intermediate rhombohedral phase, distortion of the B2 phase, known as the R-phase [6, 10, 17, 19-23]. Presence, or not, of this last phase is strongly influenced by heat treatment of the shape memory alloy [21, 22, 24]. As reported in Figure 3b, for each sample, three heat flow peaks can be observed: a single stage M → A transformation while heating and two stages A → R → M transformation while cooling. The two samples, corresponding to the 1$^{st}$ cycle and to the 40$^{th}$ cycle, present the same behaviour during heating and during cooling with slight differences. The curve corresponding to 40 cycles is slightly shifted to higher temperatures and the peaks slightly lower in comparison with those of the 1$^{st}$ cycle curve. As reported in Table 1, transformation temperatures are roughly the same for each phase transformation. But now, when the curves of the three samples are superimposed, marked differences are visible mainly for the curve corresponding to 3324 cycles. For the peak corresponding to the martensite transformation, the shift towards lower temperature is very marked, although it is not the case for the other peaks (Table 1). The height of the peaks is similar for 1 cycle and 40 cycles but strongly decreases between 40 cycles and 3324 cycles. The areas of the peaks are quite similar for the two first curves (1 and 40 cycles) but much lower for the curve corresponding to 3324 cycles. Another change is the aspect of the double peak of the austenitic transformation. For a low number of cycles, the greater peak is situated at a higher temperature while for 3324 cycles it is the contrary, indicating a change in the proportions of the two structures probably associated with the deformation mechanisms. This double peak could be interpreted as the inverse transformation of martensite to austenite with the transitory formation of phase R or to another effect. To better understand the origin of that double transformation, supplementary DSC tests are necessary.

Just after heat treatment, for the untested material, a trace of R-phase was observed on the DSC thermogram (Figure 3b). The peak of heat flow for these phase, in untested material, is very low regarding about the same phase in cycled material. This study highlights that the presence of the R-phase is due to cycling. A lot of studies reveal that temperatures of each transition phase ($M_s$, $M_f$, $A_s$ and $A_f$) and the peak temperatures ($M_p$ and $A_p$) were influenced by the heat treatment [10, 17, 19, 25, 26]. As the heat treatment, cyclic loading is now an important parameter which can disturb the phase transformation temperatures and may be at the origin of the intermediate phase (R-phase).

### 3-3 – Nanomechanical properties

For each sample, six series of 49 indentations were done and the mean values of the microhardness, the indentation modulus and creep capacity as well as the maximum depth under the load at the end of the creep time are given in the Table 2.

The standard deviation is lower than 2% for the microhardness which is a rather good result for this material with respect to its microstructure. Regarding its variation during cycling, the microhardness increases by 5.2% between 1 cycle and 40 cycles where it reaches a maximum and slightly decreases by 2.2% between 40 cycles and failure at 3324 cycles.

For the indentation modulus, the standard deviation is lower than 2%, too. Similarly to the microhardness, the modulus follows the same tendency with variations of 3.5% between 1 cycle and 40 cycles and 5.8% between 40 cycles and failure.

For the creep behaviour, the standard deviation is in the range 3.6% – 4.3%. The creep behaviour follows an inverse tendency, firstly a decrease of 1.5% followed by an increase of 2.6%. This tendency is logical.

As described in the previous paragraph, after the first unloading no residual strain was noticed, the material being completely in austenitic phase. Irreversible strain started to appear at the end of the second unloading. That means that after the $40^{th}$ cycle material was not entirely in austenitic phase, certainly some residual martensite was present. From the first cycle to the $40^{th}$ cycle, the structure of the material has changed: a small part of the austenitic phase was transformed on residual martensite. This state change imposes a microstructure change [27]. This change may explain the evolution of the mechanical properties like microhardness or indentation modulus increase and creep decrease. The same type of observation has been reported by Delobelle et al. [28]. These authors demonstrated that indentation modulus increases with grain size. They also noticed that grain orientation has an important influence on the nanomechanical properties. In Figure 4, it can be seen that the distribution of the indentation sizes (or the hardness values) is not randomly distributed but corresponds to the microstructural characteristics. In order to characterize very finely the evolution of the microstructure during cycling, a further analysis on the low cycle fatigue tests coupled with Electron Backscatter Diffraction (EBSD) is necessary.

To examine the effect of fatigue loading on TiNi instruments, Jamlech et al. [29] used the nano-indentation technique. They compared fractured instruments with new ones. Their results highlighted that the fatigue process revealed a significant decrease in the hardness and elastic modulus of the TiNi instruments. These authors concluded that the fatigue process did not result in work hardening but rather work softening. Indeed, from the first cycle to 3324 cycles (failure of the specimen) our results showed a work softening too. But if result at the $40^{th}$ cycle is taken into account, the conclusion is then modified. The TiNi shape memory alloy knows a work hardening during the first part of cycling, maybe while the stabilization of the CSS behaviour takes place [9], in this study during 40 cycles. Once the CSS behaviour stabilized, that means no evolution in the shape of hysteresis loops [12], a work softening can be observed.

**3-4 – TEM observations**

Very few studies in the TiNi memory shape alloy on the low cycle fatigue test coupled with TEM observations are reported in literature [30, 31]. In order to have a better understanding of the cyclic deformation mechanisms, TEM observations were performed after each stopped low cycle fatigue tests.
After the first cycle, the deformation microstructure is characterized by a rather low dislocation density (Figure 5) without real interaction between them. A pile-up is however visible near the top edge of the grain. Very thin microtwins, approximately 2 nm to 3 nm in width, can also be seen locally (Figure 6 a and b). Almost everywhere a high density of small precipitates can be seen, in the same manner as in the uncycled material (Figure 7). Their mean size ranges between 25 nm to 15 nm.
After 40 cycles, the dislocation density is slightly higher but the microtwins are bigger (between 50 nm and 100 nm) and more numerous (Figure 8 a and b). Furthermore, a new phenomenon has appeared, with a triple distribution of precipitate size, the first one ranging between 25 and 18 nm (in the left part of Figure 9), similar to the one observed after 1 cycle, the second one ranging between 16 and 10 nm and the third one between 6 and 2 nm (in the right part of Figure 9). The reason comes from the fact that precipitates are shorn by the mobile dislocations as shown in Figures 10 (Figure 10 a – Figure 10 b). At this stage of cycling, all the precipitates have not been shorn since dislocations are not very numerous and some areas keep the precipitates in the state of the uncycled material. In the areas where dislocations have shorn the precipitates, two distributions are seen with precipitates cut one or two times and precipitates cut several times, probably reaching the threshold of critic size before dissolution.
After failure, the dislocation density is higher than at low number of cycles, but not very high. Also the density of twins is higher, with sometimes two or even three systems in a grain (lines in Figure 11) but not everywhere. At the same time, the precipitate distribution has evolved towards a non homogeneous configuration with free precipitate channels (Figure 12). The localisation of the dislocation activity is probably responsible for the channels empty of precipitates, due to the shearing of precipitates as it has been seen at 40 cycles. Indeed the shearing of a precipitate by several dislocations will induce a size decrease of the precipitate and later its dissolution under its critical size. Locally the presence of the R phase was noted, as well as some small domains of martensite with a mean size of 200 to 300 nm.
From the first cycle to failure, it seems to exist a double and simultaneous mechanism of deformation that develops without competition: microtwinning and localisation of gliding dislocations that shear the precipitates and form channels free of them. Observations reported here are partly in agreement with those noticed in the literature. The presence of twinning has already been described by Tan et al. [32] on a tensile test. For them, phenomenon of twining appears only on the stress plateau, during the martensite transformation. Continuing deformation beyond the stress plateau, induces a detwinning of the martensite, accompanied by the production of dislocations [32]. In our study, we did not observe this detwinning. On the contrary, from the first cycle, very thin microtwins have been seen locally and then they develop up to failure. Maybe further investigations are necessary to clear up this point. On the other hand, as observed by those authors, beyond the stress plateau, there was indeed a production of dislocations. In comparison with the uncycled material, the number of dislocations slightly increased from the first cycle. As in the work of Delville et al.[30], the dislocation density slightly increases during cycling. After 40 cycles, as after failure, observations reported here revealed that dislocations shorn precipitates. This result is in agreement with that noticed by Michutta et al.[33]: precipitate size is directly governed by the density of dislocations. After failure, the precipitate distribution has evolved towards a non homogeneous configuration with free precipitate channels. The disappearance of precipitates in the channels would predominate on the increase of microtwins to induce a decrease of the nanomechanical properties as the microhardness and the indentation modulus. Several authors [13, 30, 34] published that dislocations are created during the formation of martensite in TiNi. In this study, a rather low density is observed after the first cycle. So, are dislocations formed only from the austenite to martensite regime or in the reverse transformations too? To answer this question, further investigations are necessary. For different authors [14, 35] the irreversible strain results from residual martensite. During unloading, all the martensite was not completely transformed in austenite. In this study, small domains of martensite have already been observed but only after failure. So no real relation can be established between these domains and the presence of irreversible strain. Indeed, it is between the first and the $40^{th}$ cycle that irreversible strain strongly increases. But no domain of martensite was noticed after 40 cycles. For about the same number of cycles, 10 cycles, Delville et al. [30] also found no residual martensite by TEM in the microstructure of cycled TiNi. They associated the presence of irreversible strain with an accumulation of dislocation effect. The observations reported here would rather go in this sense.

**4-Conclusion**

The aim of this study was double: in one hand, to identify and to understand deformation mechanisms which appear during cyclic loading and particularly during low cycle fatigue, and in the other hand, to explain how these deformation mechanisms can modify nanomechanical properties like microhardness or indentation modulus.
The main results may be summarized as follows:
(1)- It is clear and now acquired that the cyclic stress-strain behaviour of the TiNi SMA is composed of two stages. In the first one, during the first hundred cycles, the hysteresis loops evolve, changing their form and becoming smaller, with emergence of irreversible strain at the end of the second unloading. In the same time, two deformation mechanisms expand simultaneously and without competition: twinning and gliding dislocations. During cyclic loading, twins become bigger, and dislocations glide easily and shear precipitates. These two mechanisms imply a slight increase in the hardness and indentation modulus, but have no strong influence on the transformation temperatures. The appearance and the increase of irreversible strain, in this stage, are due to an accumulation of dislocation effect.
(2)- In the second part of the behaviour, a stabilization of the cyclic-strain-stress behaviour is observed without additional irreversible strain. Two mechanisms are always present. The density of twins and dislocations are now slightly higher. The precipitate distribution evolves towards a non homogeneous configuration due to their shearing by dislocations. This mechanism induces a micro-structure constituted of walls containing a high density of small precipitates and separated with free precipitate channels. The implementation of this structure includes the increase of twinning and by the same time a work softening and a very marked translation of the transformation temperatures towards lower temperature are observed.

**Bibliography**

[1]     C. Boller., in: M. Friswell (Eds.), Adaptative aerospace structures with smart technologies - a retrospective and future view adaptative structures: engineering applications, Inc. New-York, 2007, pp. 163-190.
[2]     F.E. Feninat, G. Laroche, M. Fiset, D. Mantovani, Adv. Eng. Mater. 4 (2002) 91-104.
[3]     K. Otsuka, X. Ren, Intermetallics. 7 (1999) 511-528.
[4]     P. Krulevitch, A.P. Lee, P.V. Ramsey, J.C. Trevino, J. Hamilton, M.A. Northrup, J. Microelectromech. Syst. 5 (1996) 270-282.
[5]     O. Doare, A. Sbarra, C. Touzé, M. Ould Moussa, Z. Moumni, Int. J. Solid. Struct. 49 (2010) 32-42.
[6]     S. Eucken, T. Duering, Acta Metal. 37 (1989) 2245-2252.
[7]     Z. Moumni, A. Vanherpen, P. Riberty, Smart Mater. Struct. 14 (2005) S287-S292.
[8]     S. Nemat-Nasser, W. Guo, Mech. Mater. 38 (2006) 463-474.
[9]     A. Paradis, P. Terriault, V. Brailovski, V. Torra, Smart Mater. Struct. 17 (2008) 1-11.
[10]    Q. Liu, X. Ma, C. Lin, Y. Wu, Mater. Sci. Eng. A. 438-440 (2006) 563-566.
[11]    M. Piedboeuf, R. Gauvin, J. Sound Vib. 214 (1998) 885-901.
[12]    A.L. Gloanec, P. Cerracchio, B. Reynier, A. Vanherpen, P. Riberty, Scripta Mater. 62 (2010) 786-789.
[13]    S. Miyazaki, T. Imai, Y. Igo, K. Otsuka, Metal. Trans. A-Phys. Metal. Mater. Sci. 17 (1986) 115-120.
[14]    C. Dunand-Châtellet, Z. Moumni, Int. J. Fatigue. 36 (2012) 163-170.
[15]    J. Shaw, S. Kyriakides, J. Mech. Phys. Solids. 43 (1995) 1243-1281.
[16]    Z. Wang, X. Zu, X. Feng, H. Mo, J. Zhou, Mater. Lett. 58 (2004) 3141-3144.
[17]    P. Filip, K. Mazanec, Scripta Metal. Mater. 32 (1995) 1375-1380.
[18]    A. Paula, J. Canejo, N. Schell, F. Braz Fernades, Nucl. Inst. Meth. Phys. Res. B. 238 (2005) 111-114.
[19]    H. Shahmir, M. Nili-Ahmadabadi, F. Naghdi, Mater. Des. 32 (2011) 365-370.
[20]    A. Miller, D. Lagoudas, Mater. Sci. Eng. A. 308 (2001) 161-175.
[21]    D. Chroback, D. Stroz, Scripta Mater. 52 (2005) 757-760.
[22]    Y. Zhou, G. Fan, Mater. Sci. Eng. A. 438-440 (2006) 602-607.
[23]    J. Kim, K. Liu, S. Miyazaki, Acta Mater. 52 (2004) 487-499.
[24]    J. Uchil, F. Braz Fernades, K. Malesh, Mater. Charact. 58 (2007) 243-248.
[25]    J. Marquez, T. Slater, F. Sczerzenie. Determining the transformation temperatures of TiNi alloys using differential scanning calorimetry. In *Proceedings of the Second International Conference on Shape Memory and Superelastic Technologies (SMST-97)*. 1997.
[26]    Y. Wang, Y. Zheng, Y. Liu, J. Alloys Compd. 477 (2009) 764-767.
[27]    J.A. Shaw, C.B. Churchill, M.A. Ladicola, Exp. Tech. 32 (2008) 55-62.
[28]    P. Delobelle, S. Dali, F. Richard, Mater. Tech. 99 (2011) 185-196.
[29]    A. Jamlech, A. Sadr, N. Nomura, Int. Endodontic J. 45 (2012) 462-468.
[30]    R. Delville, B. Malard, J. Pilch, P. Sittner, D. Schryvers, Int. J. Plasticity. 27 (2011) 282-297.
[31]    S. Kajiwara, Metal. Trans. A. 17 (1986) 1693-1702.
[32]    G. Tan, L. Yinong, P. Sittner, M. Saunders, Scripta Mater. 50 (2004) 193-198.
[33]    J. Michutta, M. Carroll, A. Yawny, C. Somsen, K. Neuking, G. Eggeler, Mater. Sci. Eng. A. 378 (2004) 152-156.

[34] T. Simon, A. Kröger, C. Somsen, A. Dlouhy, G. Eggler, Acta Mater. 58 (2010) 1850-1860.
[35] K. Gall, H.J. Maier, Acta Mater. 50 (2002) 4643-4657.

**List of figures:**

**List of tables**

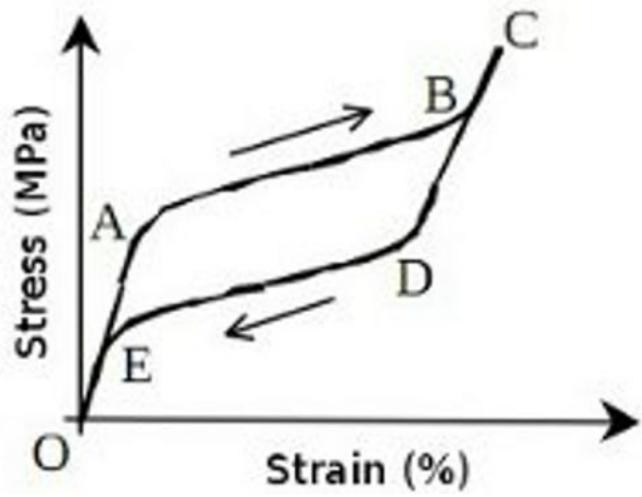

Figure1

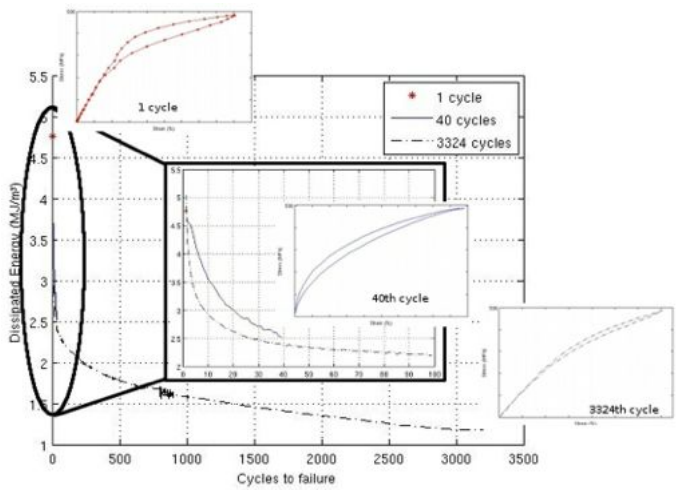

Figure2

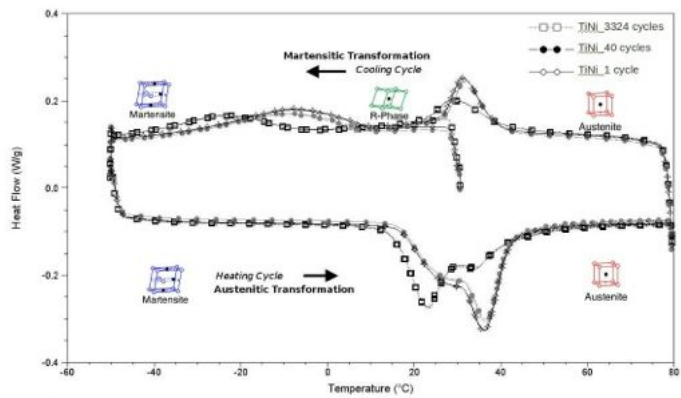

Figure3a

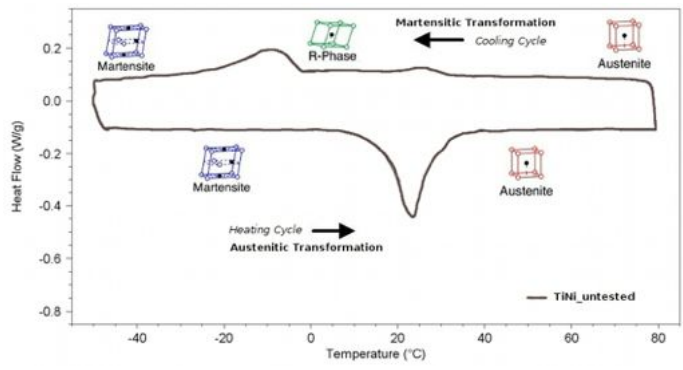

Figure 3b

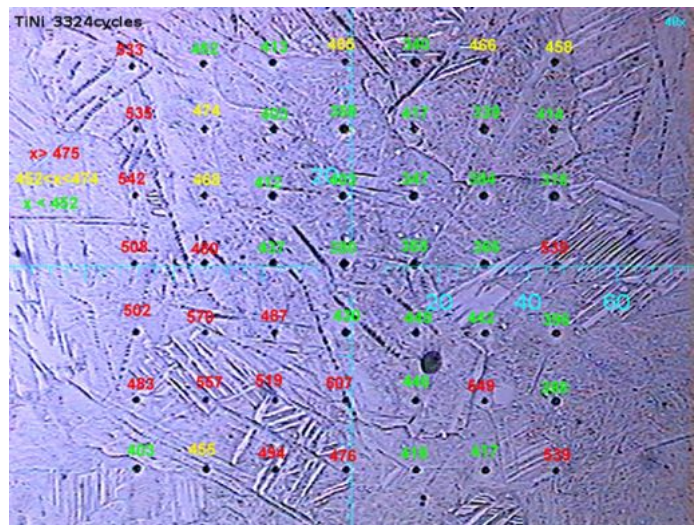

Figure4

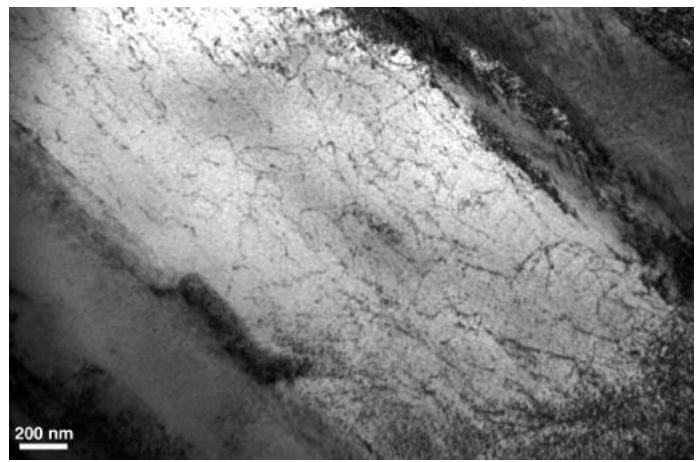

Figure5

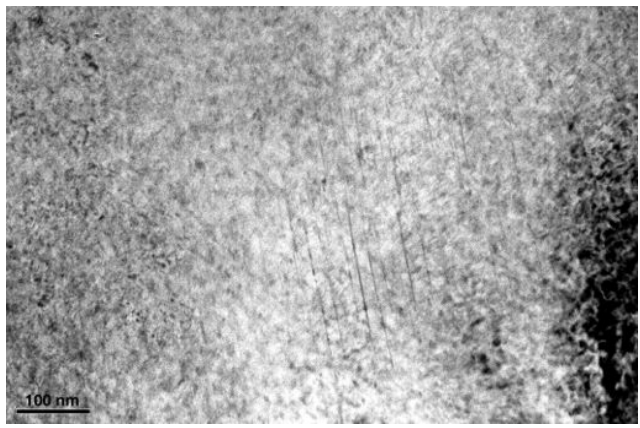
Figure6a

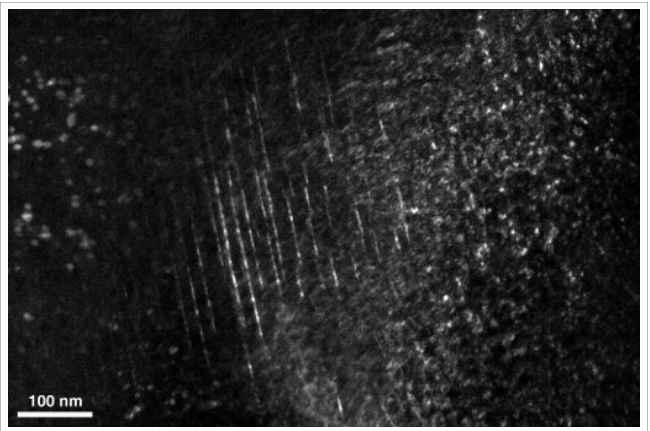
Figure6b

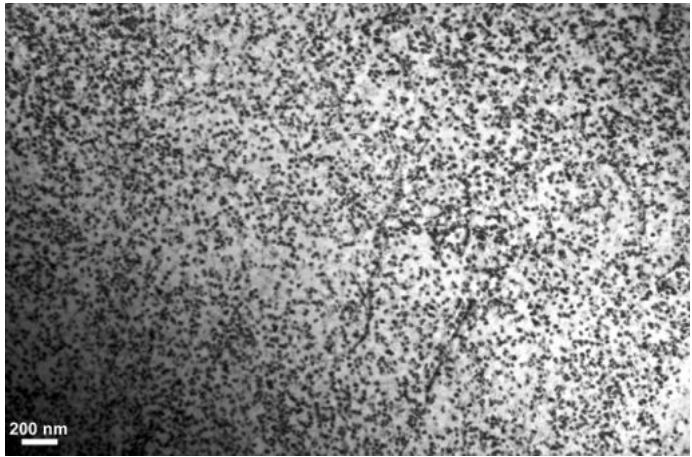
Figure7

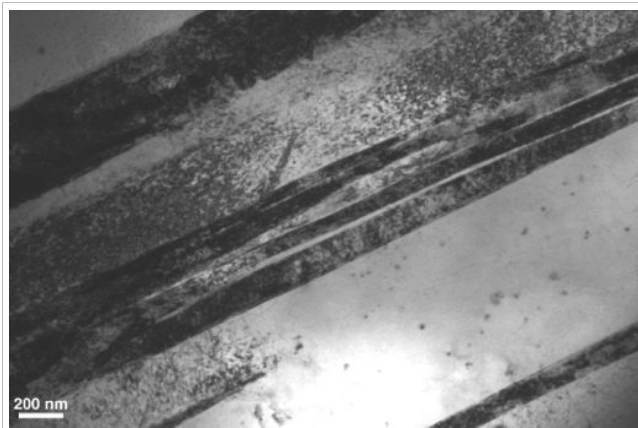
Figure8a

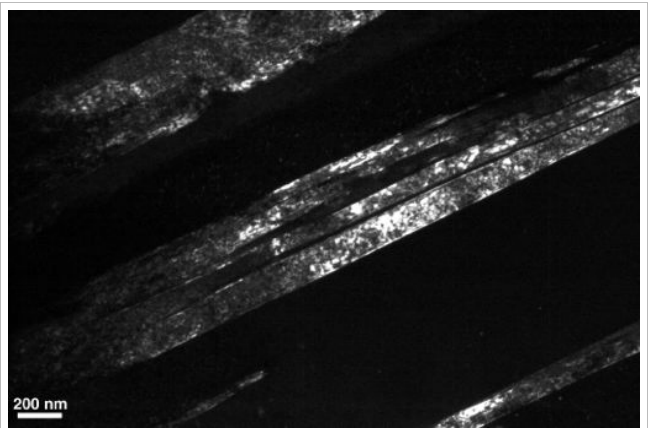
Figure8b

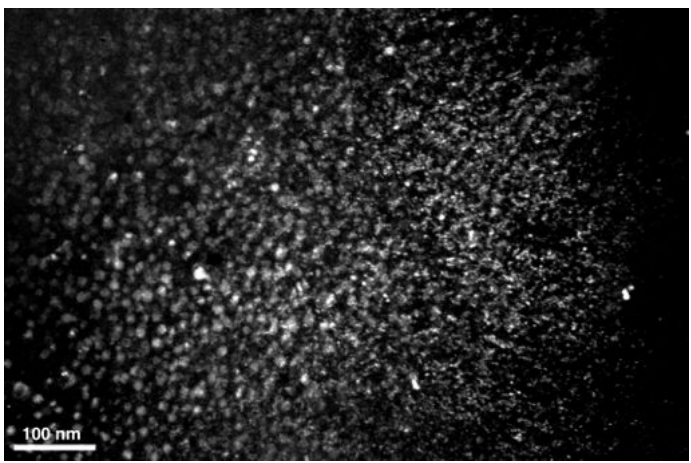

Figure9

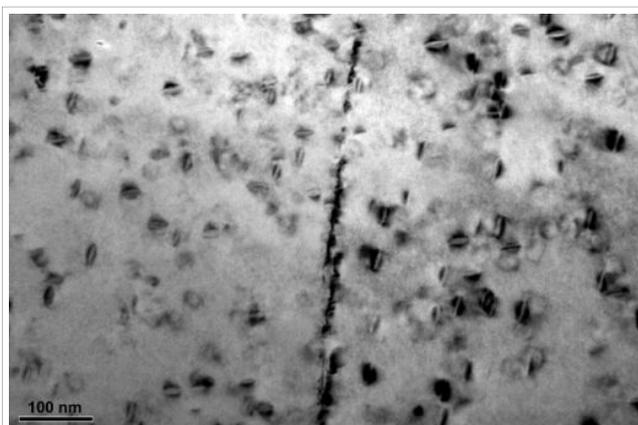
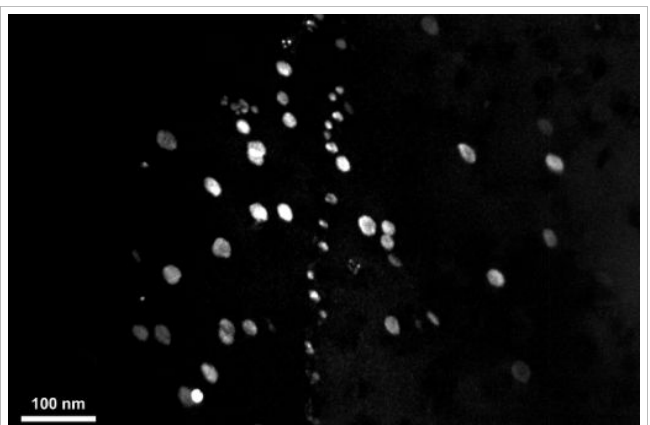

| Figure10a | Figure10b |

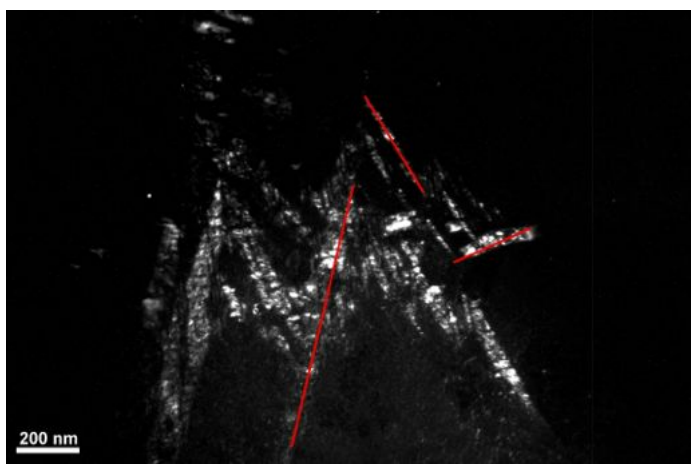

Figure11

| 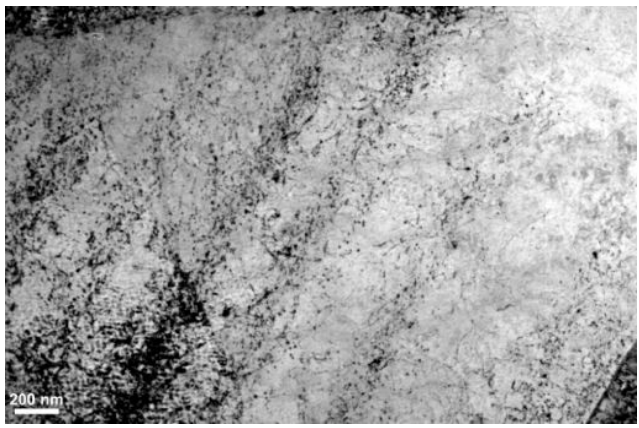 | 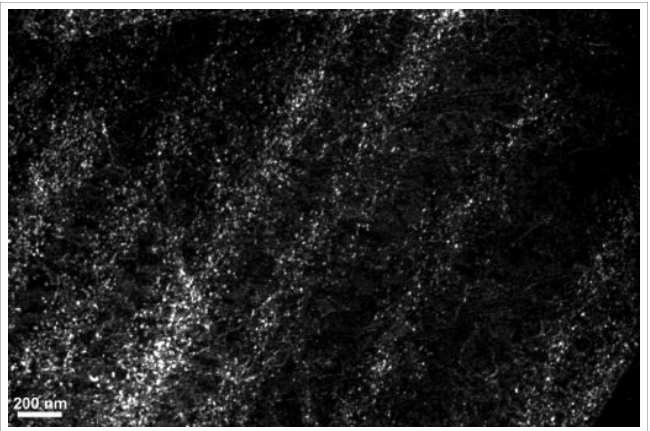 |
|---|---|
| Figure12a | Figure12b |